\documentclass[aps,prd,superscriptaddress,nofootinbib,tighten,twocolumn]{revtex4}
\usepackage{graphicx}
\usepackage{epsfig}  
\usepackage{xcolor} 
\usepackage{epsf}    
\usepackage[tbtags]{amsmath}
\usepackage{amsfonts}
\usepackage{filecontents}
\usepackage{float}
\usepackage{subfig}
\usepackage{dcolumn}
\usepackage{bm}
\usepackage{dcolumn}
\usepackage{textcomp}
\usepackage{multirow}
\usepackage{slashed}
\usepackage{float}
\restylefloat{figure}
\usepackage[]{hyperref}
  \hypersetup{
  bookmarks=true,         
  unicode=true,         
  pdftoolbar=true,     
  pdfmenubar=true,     
  pdffitwindow=true,    
  pdfstartview={FitH},    
  pdfsubject={Sterilenus@DUNE},  
  pdfnewwindow=true,     
  pdfcreator={RevTeX},
  colorlinks=true,     
  linkcolor=red,   
  citecolor=blue,    
  filecolor=black,  
  urlcolor=blue,           
  }
\usepackage{hypcap}

\newcommand{\eg}{{\it e.g.}}
\newcommand{\ie}{{\it i.e.}}

\newcommand{\beq}{\begin{equation}}
\newcommand{\eeq}{\end{equation}}
\newcommand{\beqa}{\begin{eqnarray}}
\newcommand{\eeqa}{\end{eqnarray}}

\usepackage{cancel}
\usepackage{soul}

\usepackage{paralist}
\makeatletter
\renewcommand\p@enumii{}
\makeatother
\begin{document}
\title{LSND and MiniBooNE as guideposts to understanding the muon $g-2$  results and the CDF~II  $W$ mass measurement}
\author{Waleed Abdallah}
\email[]{awaleed@sci.cu.edu.eg}
\affiliation{Department of Mathematics, Faculty of Science, Cairo University, Giza 12613, Egypt}

\author{Raj Gandhi}
\email[]{raj@hri.res.in}
\affiliation{Harish-Chandra Research Institute (A CI of the Homi Bhabha National Institute), Chhatnag Road, Jhunsi, Allahabad 211019, India}

\author{Samiran Roy}
\email[]{samroy@na.infn.it}
\affiliation{INFN - Sezione di Napoli, Complesso Univ. Monte S. Angelo, I-80126 Napoli, Italy}
\begin{abstract}
In recent times, several experiments have  observed results that are in significant conflict with the predictions of the  Standard Model (SM). Two neutrino experiments, LSND and MiniBooNE~(MB) have reported electron-like signal excesses above backgrounds. Both the Brookhaven and the Fermilab muon $g-2$ collaborations have  measured values of this parameter which, while consistent with each other, are in conflict with the SM. Recently, the CDF~II collaboration has reported a precision measurement of the $W$-boson mass that is in strong conflict  with the SM prediction. It is worthwhile to seek new physics which may underly all four anomalies.  In such a quest, the neutrino experiments could play a crucial role, because once a common solution to these anomalies is sought, LSND and MB, due to their highly restrictive requirements and  observed final states, help to greatly narrow the multiplicity of new physics possibilities that are otherwise open to the $W$ mass and muon $g-2$ discrepancies.
Pursuant to this, earlier work has shown that LSND, MB and the muon $g-2$ results can be understood in the context of a scalar extension of the SM which incorporates a second Higgs doublet and a dark sector singlet.  We show that the same model leads to a contribution to the $W$ mass which is consistent with  the recent CDF~II measurement. While the LSND, MB fits and the 
muon $g-2$ results help determine the masses of the light scalars in the model, the calculation of the oblique parameters $S$ and $T$ determines the allowed mass ranges  of the heavier pseudoscalar and the charged Higgs bosons as well as  the effective Weinberg angle and its new range.
\end{abstract}


\maketitle
\section{Introduction}
\label{sec1}
The Standard Model (SM) of particle interactions has provided us with one of the most successful theories in all of physics. Its  foundations  were laid in the 1950s and through the early 1970s\footnote{Foundational work on the SM can be considered to have begun with work in 1954 by Yang and Mills on non-abelian gauge theories~\cite{Yang:1954ek}. It was followed by Glashow's work on combining the weak and electromagnetic interactions~\cite{Glashow:1961tr} and the subsequent incorporation of the Higgs mechanism~\cite{Englert:1964et,Higgs:1964pj,Guralnik:1964eu} by Weinberg~\cite{Weinberg:1967tq} and Salam~\cite{Salam}. Crucial to subsequent theoretical and experimental progress was work by 't~Hooft and Veltman on the regularization and renormalization of spontaneously broken gauge theories~\cite{tHooft:1971akt,tHooft:1971qjg,tHooft:1972tcz}.  Finally, asymptotic freedom was shown to hold in the strong interactions by Gross, Wilczek and Politzer~\cite{Gross:1973id,Politzer:1973fx}, rounding off about two decades of remarkable developments.}.  The structure of this $SU(3)_c \times SU(2)_L \times U(1)_Y$ model of strong and electro-weak interactions  has been subsequently fleshed out and embellished by the work of thousands of experimental and theoretical physicists. It has undergone rigorous and increasingly accurate testing in numerous accelerator and non-accelerator experiments over a period of about five decades. Moreover,  these checks on its validity and precision have occurred over an impressively large range of energies. Recent examples of the breadth of its applicability are measurements of coherent elastic neutrino-nucleon scattering (CE$\nu$NS) by the COHERENT experiment~\cite{COHERENT:2017ipa}, involving keV-scale nuclear recoils on the one hand and the possible detection of a 6.3~PeV Glashow resonance event by the IceCube collaboration~\cite{IceCube:2021rpz}, on the other.  In addition, this bedrock of modern-day physics has also provided very dependable predictions for expected measurements and backgrounds for a host of experiments, which has been invaluable during their planning and data-analysis stage.

Of late, however, cracks seem to have appeared in this magnificent edifice. These worrying fissures take the form of reliable measurements which deviate significantly from its predictions. Four of the most important of these are:
\begin{itemize}
\item The  recent precise  measurement of the $W$-boson mass~\cite{CDF:2022hxs}, yielding 
$$M_W = 80,433.5 \pm 9.4~{\rm MeV}$$ using 
 8.8 fb$^{-1}$ of $\sqrt{s} = 1.96$~TeV $p\bar{p}$ collision data from the CDF~II detector at the Fermilab Tevatron. This differs from the SM prediction~\cite{Awramik:2003rn} of $$M_W^{\rm SM}  = 80, 357\pm6~{\rm MeV}$$ at a level which amounts to a discrepancy of  7$\sigma$. We also note, however, that the CDF~II measurement is in tension with the earlier measurements of $M_W$  at LEP~\cite{ALEPH:2013dgf}, Tevatron~\cite{CDF:2013dpa} and the LHC~\cite{ATLAS:2017rzl}, which lead to   the current world average~\cite{ParticleDataGroup:2020ssz} value of 
$$M_W^{\rm exp} = 80,379 \pm 12~{\rm MeV}.$$

\item The recent Fermilab muon $g - 2$ collaboration's  precise measurement of the muon anomalous magnetic moment,  $a_\mu^{\rm FNAL} = 116592040(54) \times 10^{-11}$~\cite{2104.03281}, consistent with the prior Brookhaven $g - 2$ collaboration measurement of $a_\mu^{\rm BNL} = 116592089(63) \times 10^{-11}$~\cite{Bennett:2006fi,Brown:2001mga}. These measurements, when combined,  differ from the SM theoretical prediction of $a_\mu^{\rm SM} = 116591810(43) \times 10^{-11}$~\cite{Aoyama:2020ynm,Aoyama:2012wk,Aoyama:2019ryr,Czarnecki:2002nt,Gnendiger:2013pva,Davier:2017zfy,Keshavarzi:2018mgv,Colangelo:2018mtw,Hoferichter:2019mqg,Davier:2019can,Keshavarzi:2019abf,Kurz:2014wya,Melnikov:2003xd,Masjuan:2017tvw,Colangelo:2017fiz,Hoferichter:2018kwz,Gerardin:2019vio,Bijnens:2019ghy,Colangelo:2019uex,Blum:2019ugy,Colangelo:2014qya} by a  confidence level of $4.2\sigma$.

\item The observation of unexplained electron-like excesses in the Liquid Scintillator Neutrino Detector~(LSND)~\cite{LSND:1996vlr,LSND:2001aii} at a level of $3.8\sigma$ above known SM backgrounds.

\item The observation of a similar, unexpected~$4.8\sigma$ excess in electron-like events in the MiniBooNE~(MB) detector~\cite{AguilarArevalo:2008rc, Aguilar-Arevalo:2018gpe,MiniBooNE:2020pnu}.

\end{itemize}
The CDF~II measurement has strengthened the case for new physics explanations, given the large deviation from the expected SM value. This has led to a very large number of proposed explanations involving highly diverse physics beyond the SM (BSM), including the two-Higgs doublet model (2HDM) with extensions~\cite{Fan:2022dck,Zhu:2022tpr,Lu:2022bgw,Zhu:2022scj,Song:2022xts,Bahl:2022xzi,Heo:2022dey,Babu:2022pdn,Biekotter:2022abc,Ahn:2022xeq,Han:2022juu,Arcadi:2022dmt,Ghorbani:2022vtv,Kim:2022hvh,Lee:2022gyf,Atkinson:2022qnl,Kim:2022xuo}, supersymmetry~\cite{Yang:2022gvz,Du:2022pbp,Tang:2022pxh,Athron:2022isz,Zheng:2022irz,Ghoshal:2022vzo}, triplet Higgs extensions~\cite{Cheng:2022jyi,Du:2022brr,Kanemura:2022ahw,Mondal:2022xdy,Borah:2022obi,Senjanovic:2022zwy,Bahl:2022gqg}, seesaw models~\cite{Blennow:2022yfm,Arias-Aragon:2022ats,Liu:2022jdq,Chowdhury:2022moc,Popov:2022ldh,Batra:2022pej}, models with leptoquarks~\cite{Athron:2022qpo,Cheung:2022zsb,Bhaskar:2022vgk,Chowdhury:2022dps}, analyses using SM effective theories (SMEFTs)~\cite{deBlas:2022hdk,Fan:2022yly,Bagnaschi:2022whn,Paul:2022dds,Gu:2022htv,DiLuzio:2022xns,Endo:2022kiw,Balkin:2022glu,Cirigliano:2022qdm,Borah:2022zim}, vector-like fermions~\cite{Lee:2022nqz,Kawamura:2022uft,Crivellin:2022fdf,Nagao:2022oin,Cao:2022mif}, as well as other mechanisms~\cite{Yuan:2022cpw,Strumia:2022qkt,Cacciapaglia:2022xih,Sakurai:2022hwh,Heckman:2022the,Krasnikov:2022xsi,Peli:2022ybi,Perez:2022uil,Wilson:2022gma,Zhang:2022nnh,Carpenter:2022oyg,Du:2022fqv}.

Similarly, a very large number of new physics proposals for the muon $g-2$ discrepancy have been put forth. These are too numerous for us to provide individual references here, and hence the reader is referred to the reviews in~\cite{Athron:2021iuf,Crivellin:2019mvj} for details and a full set of references. Possible solutions arise from  technicolour, supersymmetry, composite models, 2HDMs, extra dimensions, new gauge bosons, leptoquarks, Higgs triplets, vector-like leptons and very weakly interacting scalars.

 The combined significance of the LSND and MB  results is  $6.1\sigma$~\cite{Aguilar-Arevalo:2018gpe}.  The results are backed by  careful  measurements  and estimates of possible SM backgrounds in order to eliminate standard physics explanations (For discussions, see \eg~\cite{LSND:1996jxj,Katori:2020tvv,Dasgupta:2021ies,Brdar:2021cgb,Alvarez-Ruso:2021dna,MicroBooNE:2021zai} and references therein.). Many new physics explanations have been proposed for these anomalies, and for a review and references we refer the reader to~\cite{Abdallah:2022grs}. An important qualitative difference between these anomalies and those observed by the muon $g-2$ and CDF~II experiments is the multiplicity and tightness of  constraints  in the former case. This situation stems  from the low energy (and hence well tested) environment at which new physics must manifest itself to explain LSND and MB, as well as the fact that the final states in the two detectors have  been observed to be electron-like\footnote{Both detectors cannot distinguish between final-state electrons, photons or $e^+e^-$ pairs.}. It also arises because their angular and energy distributions have been measured and found to be distinctive. The constraints on LSND and MB originate from  oscillation experiments, requirements based on anomaly cancellation, a host of decay experiments, a variety of neutrino-nucleon neutral current measurements at both high and low energies, mixings between SM and sterile neutrinos, neutrino-electron scattering measurements, coherent neutrino-nucleon scattering, dark photon searches, beam dump experiments, near-detector measurements in long-baseline experiments, collider experiments,  Higgs physics, vacuum stability requirements and  electroweak precision measurements (see, for a discussion and full references~\cite{Abdallah:2020biq,Abdallah:2020vgg,Abdallah:2022grs}).
 
  Additionally, we note that most proposed new physics solutions for LSND and MB involve the tree-level {\it{in-situ}}
 production of low mass particles and lead to a restricted number of viable proposals. In contrast, many  (but not all) viable solutions to account for the muon $g-2$  and CDF~II results involve new heavy particles and loop effects, allowing a large number of new physics possibilities.
 
 There are proposals which aim to explain both the CDF~II and the muon $g-2$ measurements, as in for instance~\cite{Athron:2022qpo,Du:2022pbp,Tang:2022pxh,Lee:2022nqz,Han:2022juu,Kawamura:2022uft,Cheung:2022zsb,Nagao:2022oin,Chowdhury:2022moc,Arcadi:2022dmt,Bhaskar:2022vgk,Kim:2022hvh}. While it is not necessary, it is certainly desirable that there exist a relatively simple and natural extension of the SM which accounts for all  four of these anomalies. Due to the reasons above, when seeking a common understanding of the four anomalies, LSND and MB provide a strategically superior starting point. 
 Prior to the CDF~II W-mass result,  it was shown in~\cite{Abdallah:2020vgg} that when the SM is extended by a second Higgs doublet plus a light singlet scalar and three right-handed neutrinos, one obtains  $i)$ a resolution of both the LSND and MB anomalies, $ii)$  a portal to the dark sector, $iii)$ an agreement with the experimentally observed value of the muon $g-2$ and $iv)$  an understanding of neutrino mass via a Type I seesaw, in conformity with the observed values of neutrino mass-squared differences in oscillation experiments. Further justification for the approach adopted in~\cite{Abdallah:2020vgg}  was discussed in~\cite{Abdallah:2022grs}. In this work, we show that this model, with the parameter values used to obtain $i)$ through $iv)$, can also account for the CDF~II result on the $W$-boson mass, and, in the process, add further information on the heavier charged and neutral Higgs bosons constituting it.
 
 Finally, before  concluding this section, we take note of a fifth, significant discrepancy with SM predictions $\ie$ the $B$ anomalies, measured in observables related to the charged and neutral current decays  of the beauty quark, $b \rightarrow s l^+l^-$ and $b \rightarrow  c l^-\bar{\nu}_l$. In particular, they manifest themselves in the observable $R_K$ for the former and $R(D^{*})$ for the latter category of decays.  (For a review and references see~\cite{London:2021lfn}.). 
We do not address them in this paper. However, we note that the $R(D^{*})$ anomaly is
 amenable to a resolution by a charged Higgs in the mass range $180~{\rm GeV}\leq  m_{H^\pm} \leq 400$~GeV, as discussed recently in~\cite{Blanke:2022deg,Blanke:2022pjy,Iguro:2022uzz,Watanabe:2022ynl}.
 This mass range agrees well with the one predicted by our model in  this work (as shown below).

 Section~\ref{sec2} provides a brief description of the essential elements of the model in~\cite{Abdallah:2020vgg} and also recapitulates the important  results obtained in that paper on the LSND, MB and muon $g-2$ anomalies. Section~\ref{sec3}  provides details of the calculation of the correction to the $W$ mass which results from the model. Section~\ref{sec4}  
 summarizes and discusses our results and  section~\ref{sec5}  presents our conclusions.
 \section{Summary of the model and earlier results}
 \label{sec2}
 Our model extends the scalar sector of the SM by augmenting it with a second Higgs doublet, $\ie$, the widely studied 2HDM~\cite{Lee:1973iz, Branco:2011iw}. Additionally, it incorporates a dark singlet real scalar~$\phi_{h'}$. Three right-handed neutrinos help generate neutrino masses via the seesaw mechanism. For brevity, we have not included the neutrino part of the Lagrangian here, since it is not directly relevant to the $W$ mass issue we address in this paper. Consequently, only the scalar part of the Lagrangian is provided below. Full details of the model and the results it leads to can be found in~\cite{Abdallah:2020vgg}.

\begin{figure*}[t!]
\includegraphics[width=0.45\textwidth]{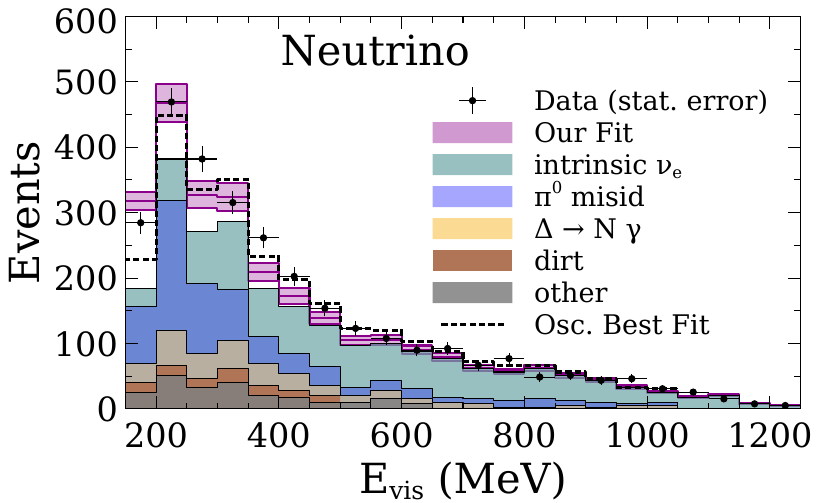}~~~~
\includegraphics[width=0.45\textwidth]{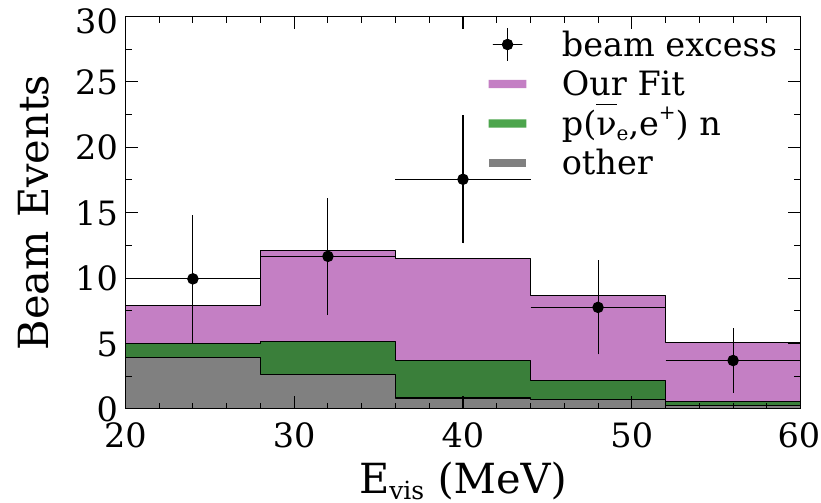}
\caption{Left panel: The MB electron-like events (backgrounds and signal), taken from~\cite{MiniBooNE:2020pnu}, versus the visible energy E$_{\rm vis}$, for neutrino beam runs. Right panel:  The distribution of the LSND events versus E$_{\rm vis}$~\cite{LSND:2001aii}.  The shaded purple region in both panels is the fit from our model, and other shaded regions are the backgrounds. (From~\cite{Abdallah:2020vgg}.).}
\label{MB-LSND-events}
\end{figure*}
\begin{figure}[t!]
\center
\includegraphics[width=0.6\textwidth]{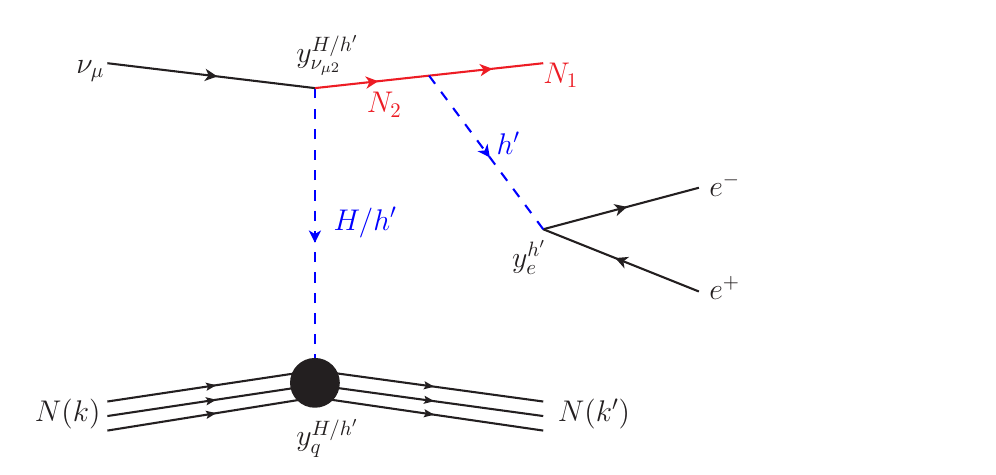}
\caption{Feynman diagram of the  scattering process in our model which leads to the excess in LSND and MB. (From~\cite{Abdallah:2020vgg}.).}
\label{FD-SP-LSND-MB}
\end{figure}

With the $\lambda_i$ denoting the usual set of quartic couplings, we  write  the scalar potential $V$ in the Higgs basis $(\phi_h,\phi_H,\phi_{h'}) $~\cite{Branco:1999fs,Davidson:2005cw} as
\begin{eqnarray}
&V\!\!=&\!\! |\phi_{h}|^2\left(\frac{\lambda_1}{2} |\phi_{h}|^2+\lambda_3 |\phi_{H}|^2+\mu_1\right) \nonumber\\
&\!\!\!+&\!\!\!\!\!\!|\phi_{H}|^2\left(\frac{\lambda_2}{2} |\phi_{H}|^2+\mu_2\right)+\lambda_4 (\phi_{h}^\dagger\phi_{H})(\phi_{H}^\dagger\phi_{h})\nonumber\\
&\!\!\!+&\!\!\!\!\!\!\phi_{h'}^2\left(\lambda'_2 \phi_{h'}^2+\lambda'_3 |\phi_{h}|^2+\lambda'_4|\phi_{H}|^2+m'\phi_{h'}+\mu'\right)\nonumber\\
&\!\!\!+&\!\!\!\!\!\!\!\bigg{[}\phi_{h}^\dagger\phi_{H}\Big{(}\frac{\lambda_5}{2} \phi_{h}^\dagger\phi_{H} +\lambda_6|\phi_{h}|^2+ {\lambda_7  |\phi_{H}|^2+ \lambda'_5 \phi_{h'}^2}\!-\!\mu_{12}\Big{)}\nonumber\\
&\!\!\!+&\!\!\!\!\!\!\phi_{h'}(m_1|\phi_{h}|^2+m_2|\phi_{H}|^2+m_{12}\phi_{h}^\dagger\phi_{H})+h.c.\bigg{]},
\end{eqnarray}
where
\begin{eqnarray}
\phi_{h}&=&\left( \begin{array}{c}
H^+_1 \\
\frac{v+H^0_1+i G^0}{\sqrt{2}} \\
\end{array} \right)\equiv\cos\beta\, \Phi_1+\sin\beta\,\Phi_2
,\\
\phi_{H}&=&\left( \begin{array}{c}
H^+_2 \\
\frac{H^0_2+i A^0}{\sqrt{2}} \\
\end{array} \right)\equiv-\sin\beta\, \Phi_1+\cos\beta\,\Phi_2,\\
\phi_{h'}&=&H^0_3/\sqrt{2}.
\end{eqnarray}
 With $v,v_i$ denoting  vacuum expectation values (vevs), so that $v^2 =v^2_1+v^2_2 \simeq  (246~{\rm GeV})^2$ and $\tan\beta=v_2/v_1$, where $\langle \Phi_i \rangle = v_i/\sqrt{2}$,  $\langle\phi_h\rangle=v$ while $\langle \phi_H\rangle\!=\!0\!=\!\langle \phi_{h'}\rangle$. Here,  $H_1^+,G^0$ are the Goldstone bosons which give the gauge bosons mass after the spontaneous breaking of electroweak  symmetry. In the basis
$\left(H_1^0, H_2^0, H_3^0\right)$, the neutral CP-even Higgs mass matrix is given by
\begin{equation} \label{CP-Even-MM}
m^2_{\cal H} = \left( 
\begin{array}{ccc}
\lambda_1 v^2 &\lambda_6 v^2  &  0\\ 
\lambda_6 v^2 & \mu_H & m_{12} v/\sqrt{2}  \\ 
 0 & m_{12} v/\sqrt{2}  &\mu_{h'}
 \end{array} 
\right), 
 \end{equation} 
where $\mu_H = \mu_2 +(\lambda_3 + \lambda_4 + \lambda_5)v^2/2$ and $\mu_{h'}=\mu' +\lambda'_3 v^2/2$.  Minimization of the scalar potential $V$ yields the  conditions,
\begin{equation}
m_1=0,~~\mu_1 = -\frac{1}{2}\lambda_1 v^2,~~\mu_{12} = \frac{1}{2}\lambda_6 v^2.
\end{equation} 
In the alignment limit ($\ie$, $\lambda_6\simeq 0$),  we diagonalize $m^2_{\cal H}$ in Eq.~(\ref{CP-Even-MM}) by $Z^{\cal H} m^2_{\cal H} (Z^{\cal H})^T = (m^{2}_{\cal H})^{\rm diag}$, with
\begin{equation}
{Z^{\cal H}}=\left(
\begin{array}{ccc}
 1 & 0 & 0 \\
 0 & \cos \delta & -\sin \delta \\
 0 & \sin \delta & \cos \delta
\end{array}
\right).
\end{equation}
Here
\begin{equation} 
{\tan{2 \delta}=\frac{\sqrt{2}\, m_{12} v}{\mu_{h'}-\mu_H},}~~~~~~~H^0_i = \sum_{j}Z_{{j i}}^{\cal H}h_{{j}}\,, 
\end{equation} 
and $i)$~$\delta$ is the scalar mixing angle between the gauge eigenstates ($H^0_2, H^0_3$) and the mass eigenstates $(H,h')$, $ii)$~$(h_1,h_2,h_3)=(h,H,h')$ are the mass eigenstates, $iii)$~$H^0_1\approx h$ is the SM-like Higgs, $m^2_h\simeq \lambda_1 v^2$ and $iv)$~$m_{12}<0$ so as to obtain $m^2_H\geq m^2_{h'}$. Therefore, the extra CP-even physical Higgs states $(H,h')$ have masses  given by
\begin{equation}
m^{2}_{H,h'}\simeq\frac{1}{2}\left[\mu_H+\mu_{h'}\!\pm\! \sqrt{(\mu_H-\mu_{h'})^2 + 2 m_{12}^2 v^2 }\right]\!.
\end{equation}
Finally, the CP-odd Higgs mass is given by
\begin{equation}
m^2_{A}=m^2_{H^\pm} +(\lambda_4 - \lambda_5)v^2/2,
\end{equation}
and the charged Higgs mass is given by
\begin{equation}
m^2_{H^\pm}=\mu_2+\lambda_3 v^2/2.
\end{equation}
After making a convenient basis rotation, the coupling strengths of the scalars $(h',H)$ with fermions (leptons and quarks)\footnote{In using these to compute the muon $g-2$ contribution, we have assumed diagonal couplings to avoid FCNCs. However, under certain assumptions, off-diagonal couplings can be used to obtain this result, $\eg$, as in~\cite{Babu:2022pdn}.} are $(y^{h'}_{f},y^{H}_{f})$. Similar considerations apply to neutrinos, where the coupling strengths of the light scalars $(h',H)$ for vertices connecting active and sterile neutrinos are $(y^{h'}_{\nu_{ij}},y^{H}_{\nu_{ij}})$ while the couplings of the scalars $(h',H)$ to the sterile states are $(\lambda^{h'}_{N_{ij}},\lambda^{H}_{N_{ij}})$. Our choice of benchmarks for the fits shown in Fig.~\ref{MB-LSND-events} is such that the values are compatible with experimental values of global fits of neutrino mass differences, as shown in~\cite{Abdallah:2020vgg}. In Fig.~\ref{MB-LSND-events}, taken from this reference,  we show our results for the fits in MB (top panels) and LSND (bottom panels).  Table~\ref{tab} collects the relevant benchmark parameters associated with the fits. 
The masses of the three right-handed neutrinos $N_i$  are also shown, since besides playing a role in giving the SM neutrinos a mass, they also participate in the interaction in LSND and MB responsible for the observed final state, see Fig.~\ref{FD-SP-LSND-MB}. This interaction involves the production of $N_2$, which decays to $N_1$ and $h' $. The latter then produces an $e^+e^-$ pair which constitutes the electron-like signal.
\begin{table}[h!]
\begin{center}
 \begin{tabular}{|@{\hspace{0.03cm}}c@{\hspace{0.03cm}}|@{\hspace{0.03cm}}c@{\hspace{0.03cm}}|@{\hspace{0.03cm}}c@{\hspace{0.03cm}}|@{\hspace{0.03cm}}c@{\hspace{0.03cm}}|@{\hspace{0.03cm}}c@{\hspace{0.03cm}}|@{\hspace{0.03cm}}c@{\hspace{0.03cm}}|}
  \hline
  $m_{N_1}$& $m_{N_2}$ & $m_{N_3}$&$y_{u}^{h'(H)}\!\!\times\!\! 10^{6}$ &$y_{e(\mu)}^{h'}\!\!\times\!\! 10^{4}$&$y_{e(\mu)}^{H}\!\!\times\!\! 10^{4}$ \\
  \hline
   85\,MeV  & $130$\,MeV & $10$\,GeV&$0.8(8)$ &$0.23(1.6)$&$2.29(15.9)$ \\
  \hline
    \hline
  $m_{h'}$& $m_{H}$ &$\sin\delta$&  $y_{d}^{h'(H)}\!\!\times\!\! 10^{6}$&$y_{\nu_{\!i 2}}^{h'(H)}\!\!\times\!\! 10^{3}$&$\lambda_{N_{\!12}}^{h'(H)}\!\!\times\!\! 10^{3}$ \\[0.05cm]
  \hline
   17\,MeV  & 750\,MeV &$0.1$& $0.8(8)$&$1.25(12.4)$&$74.6(-7.5)$ \\
  \hline
 \end{tabular}
\caption{Benchmark point used for event generation in LSND, MB and for calculating the muon $g-2$.}
\label{tab}
\end{center}
\end{table}

\begin{figure}[h!]
\centering
\includegraphics[width=0.3\textwidth]{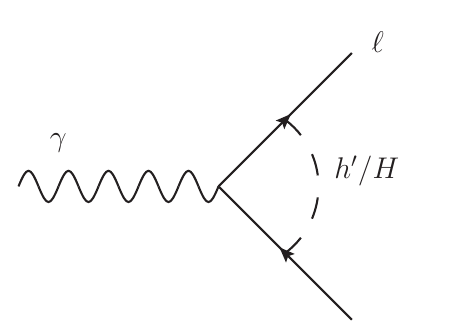}
\caption{Feynman diagram showing the one-loop contribution to $(g-2)_\mu$ from $h',H$.}
\label{FD}
\end{figure}
\begin{figure}[h!]
\centering
\includegraphics[width=0.45\textwidth]{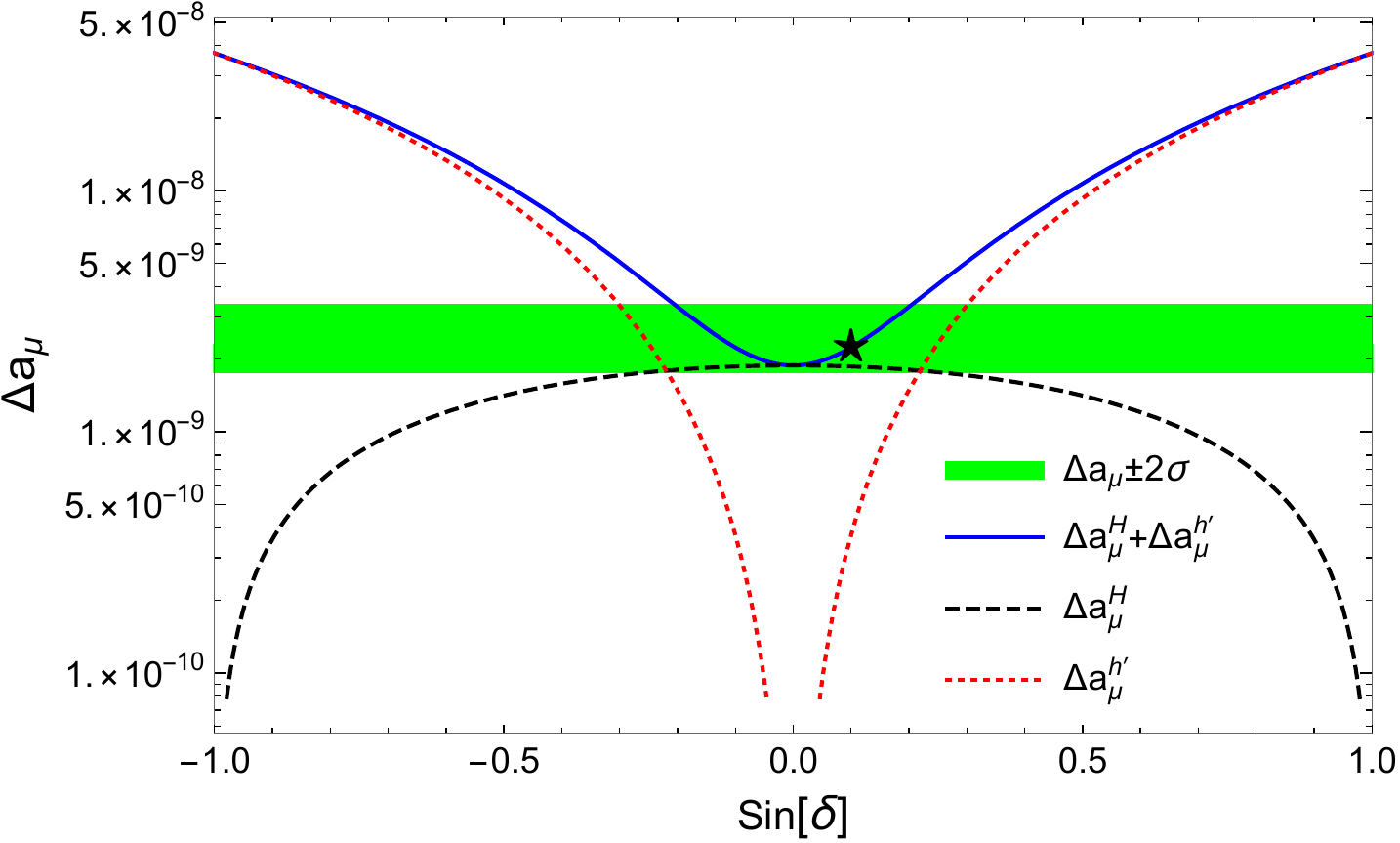}
\caption{The contribution of the light scalars $(H,h')$  to $(g-2)_\mu$ using the benchmark point shown in Table~\ref{tab} is represented by the star. The green band represents the $2\sigma$ allowed range for compatibility with the measured value of  $\Delta a_\mu$. $\sin\delta$ is a mixing between light scalars as defined in the text.}
\label{gm2}
\end{figure}

As shown in Fig.~\ref{FD}, both $h'$ and $H$ have one-loop contributions to $(g-2)_\mu$, which we have labelled as  $\Delta a_\mu$. It is can be seen from Fig.~\ref{gm2} that the $H$ contribution $\Delta a_\mu^H$ (black dashed line) is significantly larger for small values of $|\sin\delta|(\lesssim 0.2)$ while  $\Delta a_\mu^{h'} \simeq17\%$  (red dotted line) of the total (solid blue line) for our benchmark point shown in Table~\ref{tab}.

Prior to concluding this section, we note that our approach here and in~\cite{Abdallah:2020vgg} has been to find the simplest model (in the sense of Occam's razor) that provides an understanding of the anomalies while still respecting known constraints. It is completely possible that the elements of the model presented here are part of a broader new physics structure. For instance, the singlet $\phi_{h'}$ could be part of a larger dark scalar multiplet.  Its mass could receive contributions from the presence of a vev not considered here, as opposed to arising  via a mixing with the doublet scalars. It could also have both scalar as well as pseudoscalar couplings to the SM fermions. Additionally, the general 2HDM allows for a richer CP violating structure, unlike the implicit CP conserving assumptions made in our model. Exploring the feasibility and the consequences of these possibilities and their impact on the proposed solutions to the anomalies  would undoubtedly be interesting, but is outside the scope of this work.

In the next section, we calculate the contribution to the mass of the $W$-boson from this model.
\section{Calculation of the contribution to the $W$ mass in the model}
 \label{sec3}
The $W$-boson mass can be calculated in the SM using the  measured  input parameters $\{\alpha_{\rm em}, G_F, M_Z\}$ which have been accurately determined empirically~\cite{ParticleDataGroup:2020ssz}. Their central values are 
\begin{eqnarray}\label{inputs}
\alpha_{\rm em}^{-1} &=& 137.036,\nonumber \\
 G_F &=& 1.1663787 \times 10^{-5}~{\rm GeV}^{-2}, \nonumber\\
 M_{Z}&=& 91.1876~{\rm GeV}.
\end{eqnarray}
Given these values, one may calculate the $W$ mass using the relation~\cite{Hollik:1988ii},
\begin{equation}
    M_W^2 \left(1-\frac{M_W^2}{ M_Z^2}\right) = \frac{\pi \alpha_{\rm em}}{\sqrt{2} G_F}(1+\Delta r).
\end{equation}
In the above equation, $\Delta r$ represents higher order contributions, which arise in the SM (or its new physics extensions). Here $M_W$ and $M_Z$ are the renormalized masses in the on-shell scheme. One then obtains
\begin{equation}
    M_W^2 = \frac{M_Z^2}{2}\left[ 1 + \sqrt{1- \frac{4 \pi \alpha_{\rm em}}{\sqrt{2} G_F M_Z^2}(1+\Delta r(M_W^2))}\; \right].
    \label{MW}
\end{equation}
The additional radiative contribution $\Delta r$ can be expressed as~\cite{Hollik:1988ii}
\begin{equation}
    \Delta r = \Delta \alpha_{\rm em} -\frac{c_W^2}{s_W^2} \delta \rho + \Delta r_{\rm rem},
\end{equation}
where $c_W=\cos{\theta_W}, s_W=\sin{\theta_W}$ and  $\theta_W$ is the Weinberg angle. The dominant contribution to  $\Delta r$ is from two sources: $\Delta \alpha_{\rm em}$ is a  QED
correction, reflecting the evolution of  the fine structure coupling from $q^2= 0$ to $q^2 =M^2_Z$ and $\delta \rho$ reflects the vacuum polarisation  contributions to the $\rho$ parameter, whose zeroth value in the SM $(\delta\rho=0)$ is
\begin{equation}
 \rho = \frac{M_W^2}{M_Z^2c_W^2}=1.
 \end{equation}
 In the SM, such radiative contributions are largely due to the top and bottom quark  loops. Since $\Delta r$ is a function of $M_W$, Eq.~(\ref{MW}) must be solved iteratively. 

The renormalization correction to  the fine structure constant $\Delta \alpha_{\rm em}$ is  dominated by light fermions, and is 
\begin{eqnarray}
\Delta \alpha_{\rm em} &=& \frac{\alpha_{\rm em}(M^2_Z)-\alpha_{\rm em}}{\alpha_{\rm em}}\nonumber\\
&=& -\frac{\alpha_{\rm em}}{3\pi} \sum_{m_f<M_Z} Q^2_f \Bigg[\frac{5}{3}-\ln(\frac{M^2_Z}{m^2_f})\Bigg]\nonumber\\
&=& 0.05943 \pm 0.00011,
\end{eqnarray}
where $\alpha_{\rm em}(M^2_Z)=1/127.935$. $\Delta r^{[\delta \rho]} \equiv - (c_W^2/s_W^2) \delta \rho$ supplies the oblique corrections~\cite{Peskin:1991sw},  while the final term,  $\Delta r_{\rm rem}$ incorporates contributions from box and vertex diagram diagrams in the SM~\cite{Lopez-Val:2012uou}. With the input parameter values given in Eq.~(\ref{inputs}) and in addition $s_W^2 = 0.2315$, $m_t = 172.76$~GeV, $m_h = 125.25$~GeV, one finds $\Delta r^{\rm SM}=\Delta \alpha_{\rm em}+\Delta r_{\rm rem}=0.038$, and thus from an iterative solution of Eq.~(\ref{MW}), $M_W= 80.3564$~GeV. This is 7$\sigma$  and $76$~MeV below the CDF~II central value reported recently~\cite{CDF:2022hxs}. 
 
Any new physics which seeks to resolve this discrepancy must generate a $\Delta r$ which would lead to $M_W = 80.433$~GeV, the central value of the CDF~II measurement.  In our model, the singlet contribution is expected to be small (${\cal O}(\sim 1$~MeV)), as shown in~\cite{Sakurai:2022hwh}, since both its mass and its mixing with the other neutral Higgs states are small. We thus focus on calculating the correction due to the loop contributions of the additional scalars in the doublet circulating in the $W$-boson self-energy diagram,  since these dominate the correction. This dominance arises because unlike the SM, the scalar potential in the 2HDM does not respect custodial $SU(2)_{L+R}$ symmetry, and large corrections to the $\rho$ parameter are possible from the additional scalar fields ($H, H^\pm, A$). In particular, they are proportional to the mass-squared splittings among  scalars of different isospin, as apparent  in Eq.~(\ref{eq:deltarho}) below.
  
Prior to the CDF~II results, corrections to the $W$ mass for the 2HDM  have been addressed in a number of papers~\cite{Frere:1982ma,Grifols:1983gu,Grifols:1984xs,Bertolini:1985ia,Hollik:1986gg,Hollik:1987fg,Denner:1991ie,Froggatt:1991qw,Garcia:1993sb,Chankowski:1999ta,Grimus:2007if,Grimus:2008nb,Lopez-Val:2012uou,Broggio:2014mna,Hessenberger:2016atw,Hessenberger:2018xzo}. In general, the radiative correction depends on the following parameters:
\begin{equation}\label{eq:Deltar_fn}
\Delta r=\Delta r(e,M_{W},M_{Z},m_f; m_{H^i}, \sin\alpha, \tan\beta, \lambda_5)\,,
\end{equation}
where $H^i = h,H, H^\pm, A$.  $\sin\alpha$  is representative  of the mixing between the two neutral  CP-even scalars, $h,H$.  In the alignment limit for our model, $\sin(\beta-\alpha)\simeq 1$ and $\lambda_6 \simeq 0$. In general, the corrections of new scalars to the $W$-boson mass can be conveniently expressed in term of the oblique parameters $S$, $T$, and $U$~\cite{Peskin:1991sw,Eriksson:2009ws,Dugan:1991ck}, i.e.,
\begin{eqnarray}
\Delta M_W^2=\frac{\alpha_{\rm em} c_W^2 M_Z^2}{c_W^2-s_W^2}
\left[-\frac{S}{2}+c_W^2 T +\frac{c_W^2-s_W^2}{4 s_W^2}U\right].
\label{eq:mW}
\end{eqnarray}
Here $\Delta M_{W}^2 = M^2_{W} - (M_W^{\rm SM})^2$. In an effective field theory framework, the contribution from $U$ originates in dimension eight operators, and is consequently very small for most extensions to the SM. We thus assume $U \simeq 0$, and consider only the corrections from $T$ (dominant) and $S$ (subdominant) in what follows.

The dominant 2HDM contribution, in terms of a modification of the $\rho$ parameter due to an oblique correction induced by the $T$ parameter, 
calculated in the 't~Hooft-Feynman gauge  and denoted by $\delta\rho_{\rm 2HDM}$ is~\cite{Lopez-Val:2012uou} 
\begin{eqnarray}
\delta\rho_{\rm 2HDM} &=& \frac{\alpha_{\rm em}}{16\,\pi\,s_W^2\,M_{W}^2}\,
\Bigg\{\cos^2(\beta-\alpha)\left[\,F(m^2_{h},m^2_{H^{\pm}}) \right. \nonumber \\
& -&\left.F(m^2_{h},m_{A}^2)\right] + F(m^2_{A},m^2_{H^{\pm}})\nonumber \\
&+&\sin^2(\beta-\alpha)\left[\, F(m^2_{H},m^2_{H^{\pm}}) -
F(m^2_{H},m^2_{A})\right]\nonumber \\
&-&  3\cos^2(\beta-\alpha)\left[F(m^2_{H},M_{W}^2) + F(m^2_{h},M_Z^2) \right. \nonumber \\
&-& \left.  F(m^2_{H},M_Z^2) - F(m^2_{h},M_{W}^2)\right] \Bigg\}, 
\label{eq:deltarho}
\end{eqnarray}
where 
the function $F(a,b)$ is defined as follows
\begin{equation}
F(a,b) = F(b,a) = \left\{ \begin{array}{lr}
\frac{a+b}{2}- \frac{ab}{a-b}\,\log(\frac{a}{b}) & \qquad a \neq b \\
 & \\
 0 & \qquad a=b
\end{array} \right.
\label{eq:F}.
\end{equation}
Noting that  $\delta \rho= T\,\alpha_{\rm em}(M^2_Z)$, one may write, in the alignment limit ($\alpha=\beta -\pi/2$),
 \begin{align} 
    & T  =  \scriptstyle \dfrac{1}{16\pi^2 \alpha_{\rm em}(M^2_Z) v^2 } \left\lbrace  {F(m_{A}^2,m_{H^\pm}^2) +F(m_{H}^2,m_{H^\pm}^2)
    -F(m_{H}^2,m_{A}^2) } \right\rbrace, \label{eq:T}
\end{align}
In this alignment limit, it is clear that $\delta\rho_{\rm 2HDM}=\delta\rho$, where $v^2=M_W^2 s^2_W\alpha^{-1}_{\rm em}/\pi$.

Additionally, the contribution from the $S$ parameter to the $W$ mass, from Eq.~(\ref{eq:mW}) is,
 \begin{align} 
    & \Delta M_W  \simeq - \dfrac{M_{W}\,\alpha_{\rm em}(M^2_Z)S}{4(c_W^2-s_W^2)},\label{eq:MW-S}
\end{align}
where $S$ in terms of the scalars in our model and in the alignment limit is~\cite{Barbieri:2006dq}
 \begin{align} 
    & S  =  \dfrac{1}{2 \pi}\int_0^1 dx\, x(1-x)\ln{\left(\frac{x\, m_H^2+(1-x)m_A^2}{m^2_{H^{\pm}}}\right)}. \label{eq:S}
\end{align}
We have used Eqs.~(\ref{eq:deltarho})-(\ref{eq:T}) to calculate $T$, and Eq.~(\ref{eq:S}) to determine $S$ in our model. Finally, these values have been used along with Eq.~(\ref{eq:mW}) to  calculate the mass correction. In addition to the $W$ mass, changes in the oblique parameters also affect the  the effective weak  mixing angle, $\sin^2\theta_{\text{eff}}$, which has been measured in several experiments~\cite{ALEPH:2005ab}. The change (from the SM value) induced are given by~\cite{Peskin:1991sw}:
\begin{equation}
\Delta\sin^2\theta_{\text{eff}} = \frac{\alpha_{\rm em}(M^2_Z)}{c^2_W - s^2_W} \left[\frac{1}{4} S - s^2_W c^2_W T \right].  \label{s2theta}
\end{equation} 
We now proceed to present, (in the next section) results of  the  changes in the range of $S$,  $T$  and  $\sin^2\theta_{\text{eff}}$ as a consequence of the  CDF II correction to $M_W$   in our model, using the formalism above  in the next section.
\section{Discussion and Results}
 \label{sec4}
 Several recent papers have focussed on the corrections to the $W$ mass resulting from various types of 2HDM post the CDF~II measurement~\cite{Fan:2022dck,Zhu:2022tpr,Lu:2022bgw,Zhu:2022scj,Song:2022xts,Bahl:2022xzi,Heo:2022dey,Babu:2022pdn,Biekotter:2022abc,Ahn:2022xeq,Han:2022juu,Arcadi:2022dmt,Ghorbani:2022vtv,Broggio:2014mna,Kim:2022hvh,Lee:2022gyf,Benbrik:2022dja,Abouabid:2022lpg,Botella:2022rte}.
As is apparent from Eq.~(\ref{eq:deltarho}), the mass correction is governed by mass splittings among the scalars $H, H^\pm$ and $A$. Several papers have derived constraints which limit the size of these splittings and the scalar masses~\cite{Jana:2020pxx,Lee:2022gyf,Bahl:2022xzi,Abouabid:2022lpg}. Such bounds utilize the following considerations~\cite{Kanemura:1993hm}: 
\begin{itemize}
\item the boundedness of the  2HDM potential from below~\cite{Ivanov:2006yq,Branco:2011iw},
\item vacuum stability of the potential~\cite{Branco:2011iw,Barroso:2013awa},
\item  high energy perturbative unitarity 
of the amplitudes for scalar-vector, vector-vector and scalar-scalar scattering~\cite{Lee:1977eg,Kanemura:1993hm,Akeroyd:2000wc,Grinstein:2015rtl,Cacchio:2016qyh}, 
\item imposition of perturbativity on 
the Higgs quartic couplings, $\lambda_i$,~\cite{Branco:2011iw,Chang:2015goa},
\item bounds from LEP, LHC, electroweak precision data and FCNC processes~\cite{ATLAS:2018jvf,ATLAS:2018xbv,ATLAS:2018ynr,ATLAS:2020rej,CMS:2018hnq,CMS:2018nak,Chang:2015goa,Jana:2020pxx}.
\end{itemize} 
In accordance with our aim in this paper, we focus on interpreting these bounds in the context of our model, $\ie$ for the case where the second CP-even neutral boson, $H$, is much lighter than the SM Higgs $h$, which, in turn, is lighter than 
the pseudoscalar $A$ and the $H^\pm$. Applying the considerations above as discussed in~\cite{Jana:2020pxx,Lee:2022gyf,Bahl:2022xzi,Abouabid:2022lpg} to our model, we note that the lower bound on $m_{H^\pm}$ and $m_A$ is $\simeq 110$~GeV, and the upper bounds on these masses may be conservatively be assumed to be $\simeq 500$~GeV. Our results below also reflect the fact that $m_H\simeq 750$~MeV (see Table~\ref{tab}, set by fits to LSND and MB) leading to $m_{H^\pm}- m_H\approx m_{H^\pm}$ and $m_A-m_H\approx m_A$.

\begin{figure}[h!]
\center
\includegraphics[width=0.4\textwidth]{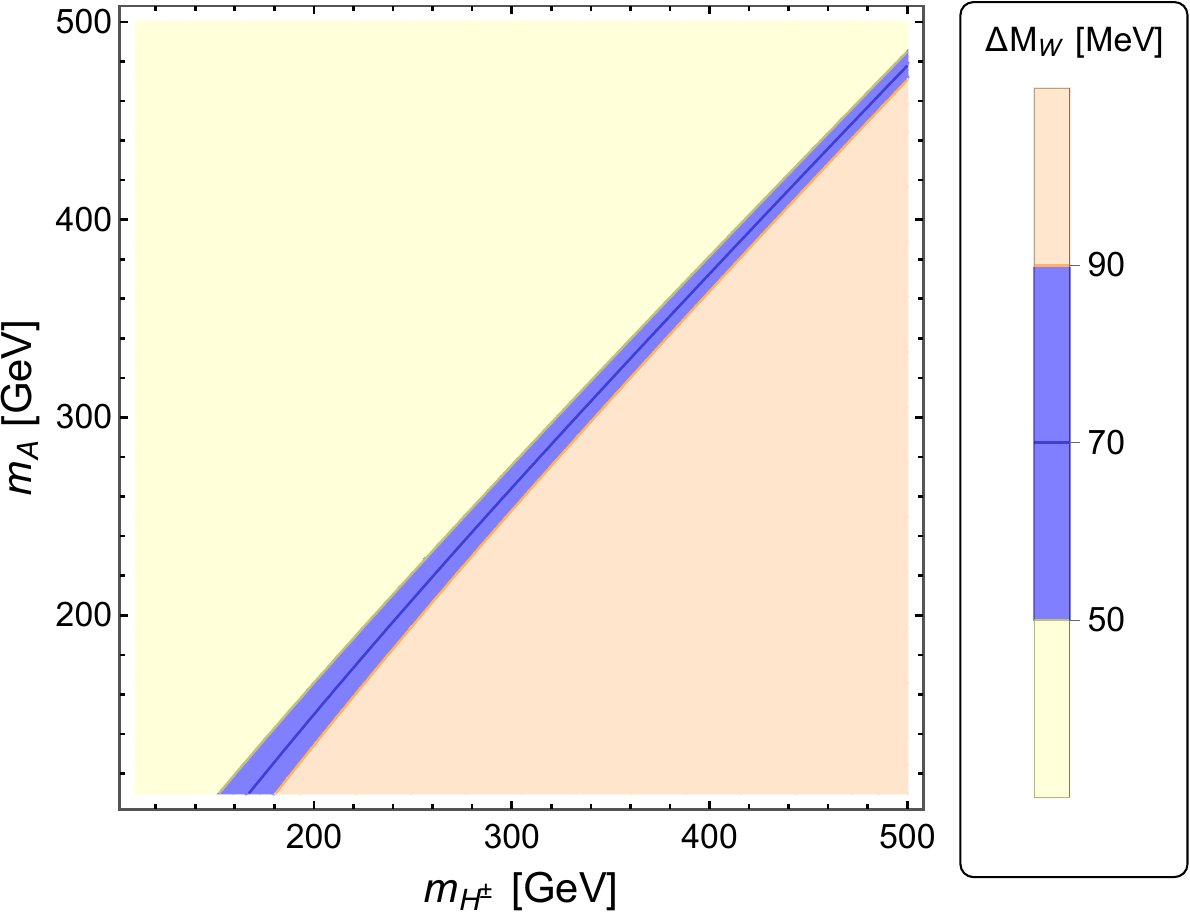}
\caption{Variation of $\Delta M_W$ in the $m_A-m_{H^\pm}$ plane. The blue band  shows the values of $m_A$ and $m_{H^\pm}$ in our model  which lead to corrections to $M_W$ in the range $50-90$~MeV. The solid blue line within the band corresponds to $\Delta M_W= 70$~MeV.} 
\label{mW-Zoom}
\end{figure}
As mentioned earlier, the CDF~II measurement and the one-loop corrected SM value differ by $\simeq 76$~MeV. Fig.~\ref{mW-Zoom} shows the blue band of values for  $m_A$ and $m_{H^\pm}$ satisfying $\Delta M_W \in [50, 90]~{\rm MeV}$. The solid blue line denotes values which yield a mass difference, $\Delta M_W$, of 70~MeV. We note that for all points in this region, $m_{H^\pm} > m_A$, and the mass differences are of the order of a couple of tens of GeV to $60$~GeV.

\begin{figure}[h!]
\center
\includegraphics[width=0.45\textwidth,height=0.28\textheight]{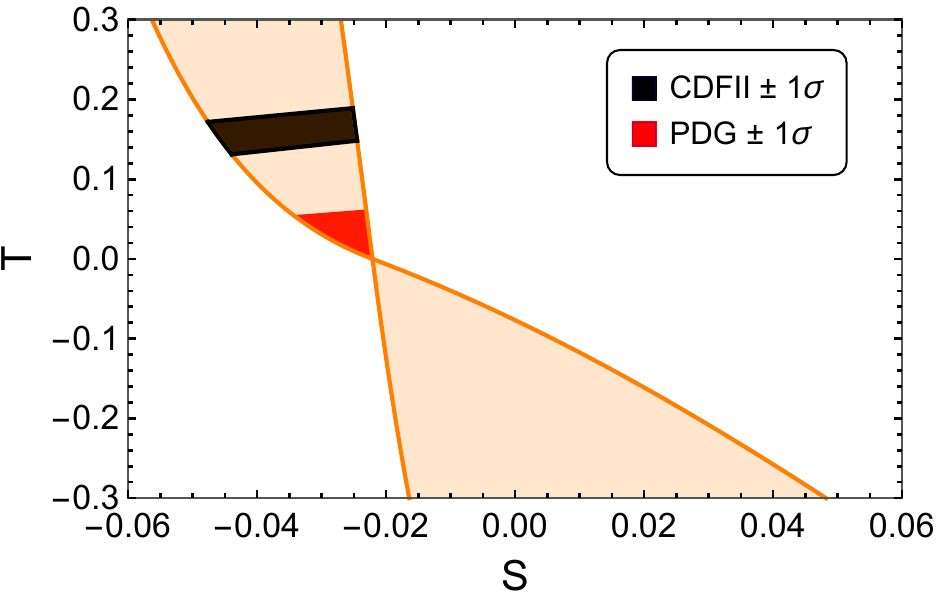}
\caption{Regions allowed by our model in the $S$ and $T$ plane.The black  band shows the $1\sigma$  region  selected by  the  $M_W$ measurement of the CDF~II experiment ($M_W=80.4335\pm 0.0094$~GeV) and the red band shows PDG data ($M_W=80.379\pm 0.012$~GeV)~\cite{ParticleDataGroup:2020ssz}.}
\label{S_T}
\end{figure}

\begin{figure}[t!]
\center
\includegraphics[width=0.45\textwidth,height=0.28\textheight]{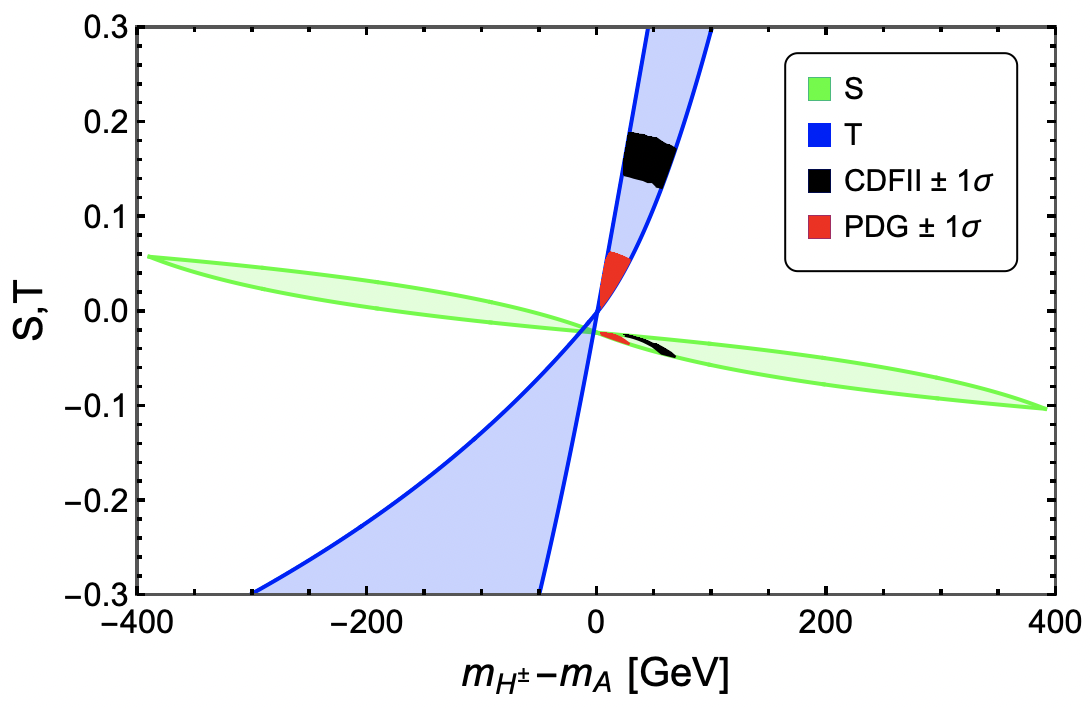}
\caption{The range of  $m_{H^\pm}-m_A$ identified by the  $1\sigma$ allowed region for $M_W$ measurement by the CDF~II experiment ($M_W=80.4335\pm 0.0094$~GeV) in our model (black regions, with $m_{H^\pm}-m_A\in[24,68]$~GeV) while the red regions (with $m_{H^\pm}-m_A\in[4,28]$~GeV) show PDG data ($M_W=80.379\pm 0.012$~GeV)~\cite{ParticleDataGroup:2020ssz}.}
\label{ST_MHP_MA}
\end{figure}
Fig.~\ref{S_T} shows the possible range of $S, T$ values when $m_{H^\pm}$  and $m_A$ are varied in the range $110$~GeV to $500$~GeV. Fig.~\ref{ST_MHP_MA} plots these parameters along with the difference between the masses of the charged Higgs and the pseudo-scalar. In both figures,  $a)$ $m_H = 750$ MeV, and $b)$ the black bands depict the region selected by the CDF measurement of the $W$ mass, while the red band shows the corresponding region for the PDG values. The latter plot identifies the mass difference  $(m_{H^\pm}-m_A)$ as  positive for the CDF results, and ranging between approximately $20$~GeV to $60$~GeV.  This differs from a more general treatment~\cite{Heo:2022dey} aimed at determining the allowed range of $S$, $T$ and mass differences between these scalars, where both a broader range as well as  positive and negative values for the difference  $(m_{H^\pm}-m_A)$ are possible  both for CDF as well as the PDG  $M_W$ mass values.

\begin{figure}[t!]
\center
\includegraphics[width=0.4\textwidth]{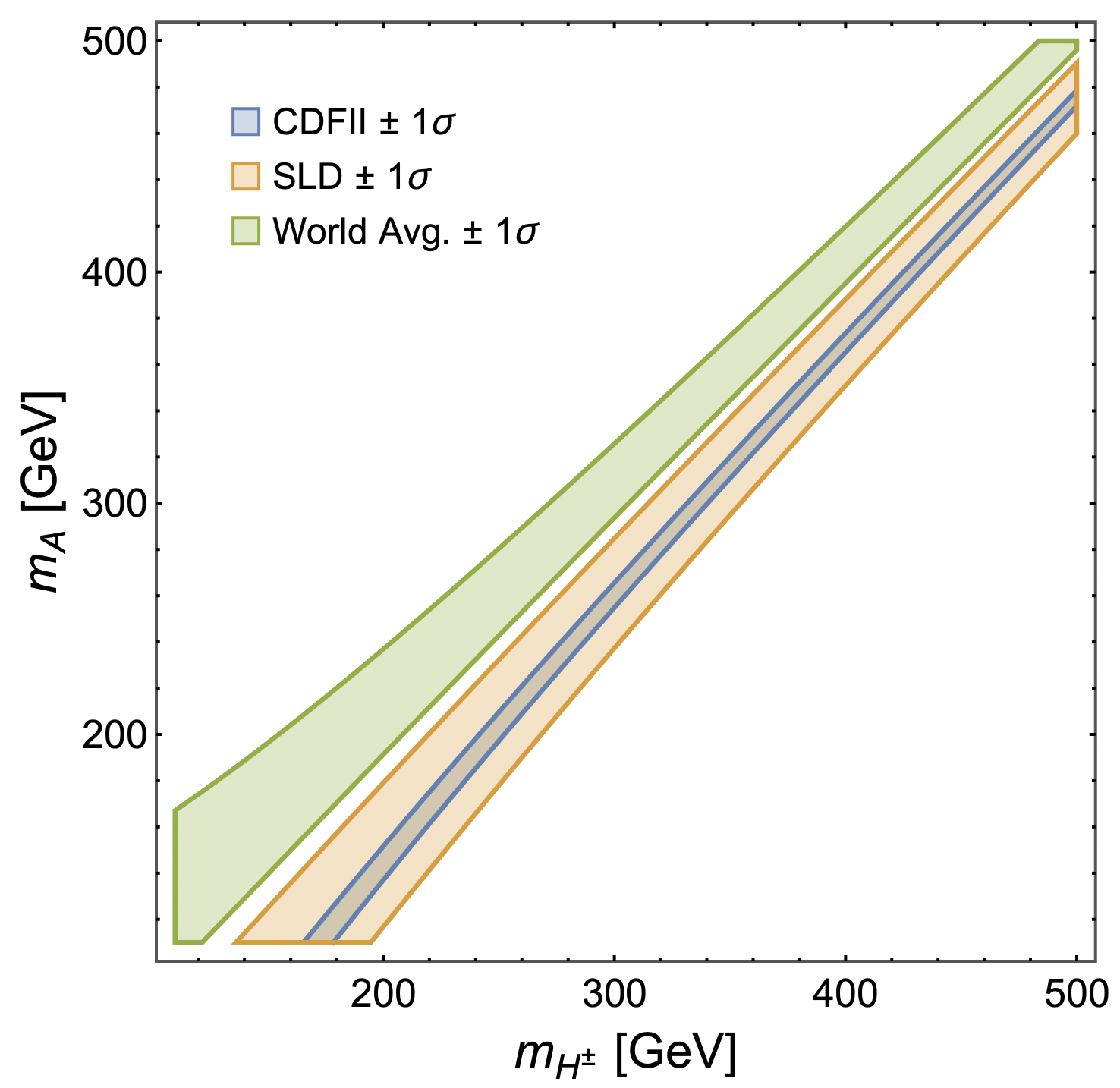}
\caption{The orange and green regions show, (in the $m_A-m_{H^\pm}$ plane) the $1\sigma$ allowed regions which lead to values compatible with experiment for $\sin^2\theta_{\text{eff}}$ in our model. They correspond to the SLD measurement ($0.2307\leq\sin^2\theta_{\text{eff}}\leq 0.23125$) (orange band) and the world average ($0.23135\leq\sin^2\theta_{\text{eff}}\leq 0.2317$) (green band). The grey band shows the $1\sigma$ allowed region for $M_W$ measurement by the CDF~II experiment ($M_W=80.4335\pm 0.0094$~GeV) in our model.}
\label{seff2}
\end{figure}
The changes in $T$ and $S$ demanded by the CDF~II mass measurement also induce a change in the value of $\sin^2\theta_{\text{eff}}$ governed by Eq.~(\ref{s2theta}). Fig.~\ref{seff2} shows the region in the $m_A-m_{H^\pm}$ plane which is in agreement with the world average measurements of this parameter~\cite{ALEPH:2005ab}, as well as those which are compatible with
the SLD measurement~\cite{ALEPH:2005ab}. These two measurements differed significantly, and from the blue-grey band in Fig.~\ref{seff2} we see that the regions favoured by 
the CDF~II measurement in our model are in tension with the world average, but in good agreement with those lying in the SLD band of values.

Our aim in this section has been  to obtain a reliable estimate of the $W$ mass correction for the model of section~\ref{sec2}. A more rigorous calculation, as in~\cite{Hessenberger:2022tcx}, would, besides all known SM higher order corrections and the one-loop new physics effects,  take into account  two-loop effects of the new scalars in the model, since they are also  expected to be sensitive to the mass splittings considered here. In addition, such a calculation would include the effects of the singlet scalar 
in our model, and not neglect its (small) mixings with the SM scalar doublets, as we have done here.

Finally, we briefly dwell on future tests of the model. An important upcoming test will be provided by the MicroBooNE experiment~\cite{MicroBooNE:2016pwy}. As part of its recent efforts~\cite{MicroBooNE:2021zai,MicroBooNE:2021nxr,MicroBooNE:2021ktl,MicroBooNE:2021pld,MicroBooNE:2021bcu} to pin down the origin of the excess seen in MB, it will search for the presence of $e^+e^-$ pairs in the final state, which are an important prediction of the model presented here. Similar searches can be conducted by the Fermilab SBN program~\cite{MicroBooNE:2015bmn}, including searches for the heavy neutral leptons that are part of the model. Invariant mass measurements for the produced pairs in these detectors  would help pin down the mass of the $h'$. We also note that the masses of $h'$ and $H$ are close to current bounds from electron beam-dump experiments like E141~\cite{Riordan:1987aw} (for $h'$) and BaBar~\cite{Lees:2014xha} (for $H$). This makes it possible for HPS~\cite{Battaglieri:2014hga} to search for $h'$ and Belle-II~\cite{Batell:2017kty} to search for $H$. Several upcoming experiments, including DUNE (see, for a discussion, $\eg$~\cite{DUNE:2020fgq,Ballett:2019bgd,Arguelles:2019xgp,Berryman:2019dme}) will be able to search for the heavy neutral leptons in the model. The heavier scalars in the model can  be searched for by the LHC. As an example, the rare (in the SM) process $pp\rightarrow t\bar{t}t\bar{t}$ is sensitive to the presence of additional scalars and pseudoscalars, via both on-shell (when the scalar mass is $> 2m_t$) and off-shell (when it is $< 2m_t$) contributions~\cite{CMS:2019rvj}.

\section{Summary and Conclusions}
 \label{sec5}
 A  large variety of new physics proposals exist for each of the four well-established  discrepancies  that have been pointed to in this paper, $\ie$ $i)$ the CDF~II $W$ mass measurement, $ii)$ the muon $g-2$ results from BNL and Fermilab, $iii)$ the LSND electron-like excess and $iv)$ the MB electron-like excess. Each of the anomalies has a significance level that demands attention and each is a serious contender for a signal of physics beyond the SM. It is possible, but unlikely, that each anomaly has its own distinct solution in new physics, not overlapping with the others. On the other hand, it is  reasonable to seek a simple, common BSM solution to these discrepancies. 
 
 Towards this end, we have noted that there is a qualitative difference between $(i)$ and $(ii)$ on the one hand, and $(iii)$ and $(iv)$ on the other. The $W$ mass  and the muon $g-2$ anomalies result from precision measurements and, if connected to new physics, are likely (but not certainly) due to higher order loop effects involving new particles and interactions. The LSND and MB observations, though anomalous, involve observed final states in detectors. While for both these experiments, the final states could be either electrons, photons or $e^+e^-$ pairs (given the inability of the two detectors to distinguish between these possibilities), this still limits the number of new physics possibilities compared to the $W$ mass  and the muon $g-2$ anomalies. Further restrictions are provided by the requirement that the observed angular and energy distributions for both experiments must be fit by the new physics. Consequently, in  attempts to find a common resolution, the LSND and MB anomalies provide a superior starting point that can usefully narrow the solution space considerably for the other two anomalies.

 Earlier work~\cite{Abdallah:2020vgg}, summarised here,  has shown that 
a 2HDM with a light singlet scalar and three right-handed neutrinos provides very good fits to LSND and MB data, as well as account for the measured muon $g-2$ value and global data on neutrino mass squared differences.  In this work we have shown that the model is also capable of providing a correction to the mass of the $W$-boson that brings it in conformity with the recent CDF~II result. While remaining natural and economical, the 2HDM contains enough elements so that both LSND and MB, which are primarily effects 
at low (MeV-GeV) energies can be bridged and connected with the CDF~II result, which requires large mass splittings and heavier scalars (few hundred GeV).

The solution  of the anomalies $(ii)$ to $(iv)$ helps obtain reference mass values for the two additional CP-even scalars in the model, $m_{h^{\prime}} \simeq 17$~MeV, which is very nearly a
 singlet, and $m_H\simeq 750$~MeV, which is a member of the second doublet. The third CP-even scalar is very closely aligned to the SM Higgs in mass and couplings. Table~\ref{tab} provides a summary of the additional CP-even scalars  and  right-handed neutrino reference masses, as well as some of their important couplings to fermions. These values emerge from the fits to LSND and  MB as well as the calculation of the correction to the muon $g-2$ in the model. Overall, these help pin down reference values for the low-mass particles and their couplings in the model.

Addressing the CDF~II anomaly  provides additional mass information on the  other (heavier)  scalars in the model, $\ie$ the pseudoscalar $A$ and the charged Higgs~$H^\pm$. The solution space picked out demands that $m_{H^\pm} > m_A$, with masses of both below $500$~GeV, and a mass difference between them of $20$~GeV to $60$~GeV. We note that these results pertaining to our model are in consonance and agreement with recent work seeking to understand the $W$ mass result within the context of 2HDMs in a more general setting~\cite{Fan:2022dck,Zhu:2022tpr,Lu:2022bgw,Zhu:2022scj,Song:2022xts,Bahl:2022xzi,Heo:2022dey,Babu:2022pdn,Biekotter:2022abc,Ahn:2022xeq,Han:2022juu,Arcadi:2022dmt,Ghorbani:2022vtv,Broggio:2014mna,Kim:2022hvh,Lee:2022gyf}. 

It is hoped that the considerations in this work help narrow the search for new physics by pointing to a candidate model which is both a simple and natural extension of the SM, as well as being one that provides a possible explanation for important outstanding anomalies.

\section*{Acknowledgements}
 RG is grateful to William Louis for his help with our many questions on LSND, MB and MicroBooNE. He would also like to acknowledge helpful discussions with Joan Sol\`a, and 
  express appreciation for the many interesting conversations he has had with Boris Kayser and Geralyn Zeller on the anomalies which are the subject of this paper. 
 
WA, RG and SR also acknowledge support from the XII Plan Neutrino Project of the Department of Atomic Energy and the High Performance Cluster Facility at HRI (http://www.hri.res.in/cluster/).
\bibliographystyle{apsrev}
\bibliography{NU-bib}
\end{document}